\documentclass[pra,aps,twocolumn,superscriptaddress,nofootinbib,showpacs]{revtex4}

\usepackage{latexsym}
\usepackage{epsfig}
\usepackage{graphics}
\usepackage{amsmath}
\usepackage{xspace}
\usepackage{amssymb}
\usepackage{amsthm}

\theoremstyle{plain}
\newtheorem{theorem}{Theorem}
\newtheorem{lemma}{Lemma}

\theoremstyle{definition}

\theoremstyle{remark}

\usepackage{hyperref}
\begin{document}

\title{Simulating Hamiltonian dynamics using many-qudit Hamiltonians
and local unitary control}

\author{Michael J. Bremner} \email{bremner@physics.uq.edu.au}
\affiliation{School of Physical Sciences, The University of
  Queensland, Queensland 4072, Australia} \affiliation{Institute for
  Quantum Information, California Institute of Technology, Pasadena CA
  91125, USA}

\author{Dave Bacon}
\email{dabacon@cs.caltech.edu} \affiliation{Institute for Quantum Information,
California Institute of Technology, Pasadena CA 91125, USA}
\affiliation{Department of Physics, California Institute of Technology,
Pasadena CA 91125, USA}

\author{Michael A. Nielsen} \email{nielsen@physics.uq.edu.au}
\affiliation{School of Physical Sciences, The University of
  Queensland, Queensland 4072, Australia} \affiliation{School of
  Information Technology and Electrical Engineering, The University of
  Queensland, Queensland 4072, Australia} \affiliation{Institute for
  Quantum Information, California Institute of Technology, Pasadena CA
  91125, USA}

\begin{abstract}
  When can a quantum system of finite dimension be used to simulate
  another quantum system of finite dimension?  What restricts the
  capacity of one system to simulate another?  In this paper we
  complete the program of studying what simulations can be done with
  entangling many-qudit Hamiltonians and local unitary control.  By
  \emph{entangling} we mean that every qudit is coupled to every other
  qudit, at least indirectly. We demonstrate that the only class of
  finite-dimensional entangling Hamiltonians that aren't universal for
  simulation is the class of entangling Hamiltonians on \emph{qubits}
  whose Pauli operator expansion contains only terms coupling an
  \emph{odd} number of systems, as identified by Bremner {\it et.
    al.}  [Phys. Rev.  A, {\bf 69}, 012313 (2004)].  We show that in
  all other cases entangling many-qudit Hamiltonians are universal for
  simulation.
\end{abstract}
\maketitle

%
%
\section{Introduction}

\subsection{Overview}

%
%
One remarkable aspect of Nature is that it can be modeled by
equations whose solution may be obtained by algorithmic means.  This
empirically observed fact allows us to construct physical theories
that make predictions as to how Nature will behave. Of course, while
we can simulate Nature, our capacity to do so is limited by the way we
choose to perform the simulation and the complexity of the system to
be simulated.  Feynman's landmark paper on quantum
computation~\cite{Feynman82a} discussed the apparent inability of
classical computers to efficiently simulate quantum systems and
suggested that a quantum computer might succeed where classical
computers fail.  In this paper we study a class of simulation
protocols motivated by the example of quantum computation. In
particular, we examine the following question: given a composite
system with a finite-dimensional Hamiltonian and the ability to
perform arbitrary local unitary operations, what other Hamiltonians
can we simulate?

%
%
The simulation of quantum systems by quantum computers is a topic that
has attracted considerable attention. A considerable literature
(see~\cite{Somma02a} and references therein) addresses the question of
how to adequately simulate physically interesting closed quantum
systems.  Issues of particular interest include the complexity of
protocols for simulating initial states, simulating evolutions, and
for extracting physically important information from the final state
of the computer.  Each of these issues must be addressed in any
comparative study of quantum and classical computers, and their
capacity to simulate Nature.

While state preparation and measurement are vital elements of any
simulation of a quantum system, we focus in this paper on the
simulation of \emph{evolutions} of systems. Hamiltonian simulation
protocols using single-qudit unitary operations as an additional
resource have received considerable attention in recent years due to
their relationship with various models of quantum computation.  One of
the more noteworthy advances was the discovery that all two-body
Hamiltonians can simulate all other Hamiltonians on the set of qudits
that they entangle, when combined with single-qudit unitary
operations~\cite{Dodd02a, Wocjan02a, Bennett01a, Leung01a, Dur01a,
  Nielsen01c, Wocjan02b, Wocjan02c, Vidal01b, Vidal01c}. This body of
work also demonstrated that these Hamiltonians could be used to
\emph{efficiently} simulate any other two-body Hamiltonian that acts
on the network of qudits they entangle. This includes a Hamiltonian
that can implement the {\sc cnot} operation, thus implying that all
entangling two-body Hamiltonians and single-qubit unitary operations
are universal for quantum computation.

The results and tools used to study two-body Hamiltonian simulation
have been applied fruitfully to several related problems.  There is
now a considerable literature on time-optimal strategies for
simulating two-qubit Hamiltonians and quantum gates; see, for example,
\cite{Khaneja01a,Bennett01a, Khaneja02a, Bremner02a,Vidal02a,
  Hammerer02a,Bullock02a,Zhang03a,Zhang03b,Haselgrove03a,Childs03a,Schende03a,Schende03b,Vidal04a,Vidal04a,Vatan04a,Zeier04a},
and references therein.  This body of investigation has led to
interest in applying these theoretical results to practical proposals
for quantum computation~\cite{Hill03a}.

More recently, studies have focused on using systems with many-qubit
interactions for Hamiltonian simulation and gate-synthesis
\cite{Bremner03a,Haselgrove03a,Zeier04a,Bullock03a,Bullock03b}. A
number of these papers have investigated the structure of systems with
many-body interactions for the purposes of gate synthesis and
algorithm design \cite{Haselgrove03a,Zeier04a,Bullock03a,Bullock03b}.
Several authors have recently examined the effects of many-body
interactions in quantum dot~\cite{Mizel04a} and optical
lattice~\cite{Pachos04a, Pachos04b} systems.

For the purposes of this paper we are most concerned with the work
in~\cite{Bremner03a}, where the authors established which Hamiltonians
with many-qubit interactions are universal when combined with the
ability to perform arbitrary single-qubit unitary operations. In a
similar vein we examine which Hamiltonians with many-\emph{qudit}
interactions are universal when combined with arbitrary
single-\emph{qudit} unitary operations.  Our final result is a
striking generalization of the conclusion in~\cite{Bremner03a}.
\cite{Bremner03a} showed that the only class of non-universal
entangling Hamiltonians on qubits are the \emph{odd} entangling
Hamiltonians, i.e., those Hamiltonians whose Pauli operator expansion
contains only terms coupling an odd number of qubits.  Furthermore,
\cite{Bremner03a} showed that the odd entangling Hamiltonians can all
simulate one another, so there is a sense in which there are only two
types of many-qubit entangling Hamiltonian.  Remarkably, in this paper
we will see that when the systems involved are not all qubits, this
structure actually \emph{simplifies}, with \emph{all} entangling
Hamiltonians capable of simulating all other entangling Hamiltonians,
i.e., we show that apart from the many-\emph{qubit} case, there is
only one type of many-body entangling Hamiltonian.

Our primary concern in this paper is with questions of universality in
many-qudit systems, without regard to the issue of complexity. Thus,
when we say a set of resources is \emph{universal} on a set of qudits,
we are stating that these resources can be used to simulate any
Hamiltonian on those qudits, without implication that this simulation
is efficient or inefficient. This is in contrast to the notion of
\emph{universality for quantum computation} which requires that any
universal set of resources can simulate a standard gate set with a
polynomial overhead in the number of qubits used. That said, it is
often possible to exploit the structure of certain classes of
many-body Hamiltonians to develop efficient simulation algorithms. For
instance much headway can be made in developing efficient Hamiltonian
simulation protocols that use $k$-local Hamiltonians by adapting the
methods developed for systems of qubits in~\cite{Bremner03a} to
systems of qudits.

\subsection{Terminology and statement of results}

%
%
Before turning to the discussion and proof of the main results of this
paper it is helpful to introduce some terminology. Generally, we will
use the term \emph{qudit} to describe \emph{any} quantum system with a
finite-dimensional state space.  As an example of our usage, a
three-qudit system might contain a two-dimensional system (a qubit), a
five-dimensional system, and a four-dimensional system.

We are interested in the properties of the Hamiltonian dynamics of an
$n$-qudit system. As we will see, a great deal can be said about the
properties of a Hamiltonian simply by examining its structure in a
suitable representation. In~\cite{Bremner03a}, the authors found that
the universality properties of a many-body Hamiltonian acting on
qubits could be identified by expanding it in the Pauli-operator
basis, i.e., tensor products of $X$, $Y$, $Z$ and $I$. In this paper
we expand upon this analysis by examining the properties of an
$n$-qudit Hamiltonian written in a $d$-dimensional generalization of
the Pauli basis.

An arbitrary Hamiltonian on $n$ qudits can be uniquely written as
\begin{equation}
\label{H} H= \sum_\alpha h_\alpha H_\alpha,
\end{equation}
where $h_\alpha$ are real coefficients and each $H_\alpha$ in the
expansion is a tensor product of Hermitian operators acting on the
individual quits,
\begin{equation}
\label{Halpha} H_\alpha = \bigotimes_{j=1}^{n}H_{\alpha}^{(j)},
\end{equation}
where $H_\alpha^{(j)}$ acts on qudit $j$, and is either the
identity operator, or one of a set of traceless Hermitian matrices
known as the \emph{Gell-Mann matrices}.  The Gell-Mann matrices
generalize the Pauli matrices, and thus this expansion is a
generalization of the expansion for qubits used
in~\cite{Bremner03a}.  The Gell-Mann matrices for a
$d$-dimensional quantum system consist of: (a) $d-1$ matrices of
the form
\begin{equation}
\label{eq:W_m} W_m = \frac{1}{\sqrt{m(m-1)}}\left(\sum_{b=1}^{m-1}
|b\rangle\langle b|-(m-1)|m\rangle\langle m|\right),
\end{equation}
where $2\leq m \leq d$; and (b) the Pauli-like matrices:
\begin{eqnarray}
X_{ab} & = &\frac{1}{\sqrt{2}}(|a\rangle\langle b| +
|b\rangle\langle a|) \\
Y_{ab} & = &\frac{-i}{\sqrt{2}}(|a\rangle\langle b| -
|b\rangle\langle a|)
\end{eqnarray}
where $1\leq a<b\leq d$. These act as the Pauli $X$ and $Y$ on the
two-dimensional subspace spanned by the vectors $|a\rangle$ and
$|b\rangle$.  We sometimes refer to the $W_m$ matrices as \emph{Cartan
  subalgebra} elements of the Gell-Mann matrices, since they span a
Cartan subalgebra of the Lie algebra $su(d)$ generated by the Gell-Mann
matrices.  However, it is worth emphasizing that we do not use any special
properties of Cartan subalgebras, and the reader does not need to be familiar
with the properties of Cartan subalgebras to follow the details of the paper;
our use of the term is a convenience of nomenclature only.  Note that the
Gell-Mann matrices are traceless and Hermitian, and form a complete basis for
traceless Hermitian matrices.

The representation Eq.~(\ref{H}) is useful as it highlights which
qudits interact and which do not.  In particular, given a term
$H_\alpha$ let $S_\alpha$ be the set of qudits upon which $H_\alpha$
acts non-trivially, that is, the set of qudits for which
$H_{\alpha}^{(j)}$ is traceless. We say that the qudits in $S_\alpha$
are \emph{coupled} by $H_\alpha$ and refer to $H_\alpha$ as a
\emph{coupling term}. We also say that $H_\alpha$ is \emph{entangling}
on the set $S_\alpha$. More generally, we say that a Hamiltonian $H$
is \emph{entangling} on some set of qudits if it is not possible to
partition this set of qudits into two non-trivial sets $S$ and
$\bar{S}$ such that every term $H_\alpha$ in the expansion of $H$
couples either a subset of $S$ or of $\bar{S}$. In graph-theoretic
language, if the qudits corresponded to vertices on a hypergraph and
the couplings corresponded to hyperedges, the condition that the
Hamiltonian is entangling on a set of qudits is simply that the
hypergraph connects this set. As such we say that a Hamiltonian
\emph{connects} the set of qudits it entangles.

Our strategy for demonstrating universality in this paper is to show
that some set of resources is capable of simulating another set
already known to be universal. In particular, we make reference to two
theorems that categorize large classes of Hamiltonians as universal up
to single-qudit unitary operations. The first was mentioned in the
introduction: two-body entangling Hamiltonians are universal for
quantum computation~\cite{Wocjan02c}.  Using the terminology just
introduced, this theorem may be stated as follows~\cite{Wocjan02c}:
\begin{theorem}
\label{thm:2local} Suppose $H$ is a two-body Hamiltonian, that
is, every coupling term in the Gell-Mann expansion of $H$ couples at
most two qudits. If $H$ is entangling on a set of $n$ qudits (that is,
the coupling terms in $H$ connect these qudits) then evolutions of $H$
together with single-qudit unitary operations are universal for
quantum computation on these $n$ qudits.
\end{theorem}
The second universality theorem that we use involves Hamiltonians
acting on sets of qubits that have coupling terms that may couple more
than two qubits. This theorem is stated \cite{Bremner03a}:
\begin{theorem}
\label{thm:nqubituniv} Suppose $H$ is an arbitary entangling Hamiltonian
on a set of $n$ qubits. Evolutions of $H$ and single-qubit unitary
operations are universal on those $n$ qubits if and only if the
Gell-Mann (i.e., Pauli) expansion of $H$ contains at least one
coupling term that couples an even number of qubits.
\end{theorem}

Theorem \ref{thm:nqubituniv} tells us that for a Hamiltonian acting on
qubits alone to be universal, it must have a coupling term acting on an
even number of qubits. If $H$ does not contain such a coupling term
then we shall call it an \emph{odd} Hamiltonian, since all its terms
couple an odd number of qubits.  What can the odd Hamiltonians
simulate? This question was also answered in~\cite{Bremner03a}:
\begin{theorem}
\label{thm:oddhams} Let $H$ be an odd entangling Hamiltonian, that is,
every term in the Gell-Mann expansion of $H$ couples an odd number of
qubits.  Then $H$ and single-qubit unitaries can simulate any other
odd Hamiltonian on the $n$ qubits.
\end{theorem}
\cite{Bremner03a} also demonstrated that the Lie algebra generated by
the odd entangling Hamiltonians on $n$ qubits (and local unitaries)
corresponds to the Lie algebras $so(2^n)$ and $sp(2^n)$, for even or
odd $n$ respectively.  Furthermore,~\cite{Bremner03a} showed that the
odd Hamiltonians can be made universal with appropriate encodings.

%
In this paper we demonstrate that if a Hamiltonian is entangling on a
set of qudits, then this Hamiltonian is universal on those qubits,
when assisted by local unitary operations. The only exception to this
result is the special case when the Hamiltonian is an odd Hamiltonian
acting on qubits only.

\subsection{Outline}
\label{sec:outline}

%
%
Theorem~\ref{thm:2local} shows that if a Hamiltonian connects a set of
qudits with two-qudit couplings then this Hamiltonian is universal
with single-qudit unitaries. Our strategy in this paper is to show
that a many-body Hamiltonian (that isn't one of the odd qubit-only
Hamiltonians) connecting a set of $n$ qudits can simulate a two-body
Hamiltonian connecting the same set of qudits.  This is done by
defining a series of simulation protocols, each identifying broad
classes of Hamiltonians that any entangling Hamiltonian can simulate,
until we arrive at the eventual result.

The structure of the paper is as follows.  In
Section~\ref{sec:simple_sims} we introduce some simple general
simulation techniques that are used often in this paper.
Section~\ref{sec:term_isolation} introduces a simulation technique
known as \emph{term isolation}.  This simulation technique allows us
to simulate any particular coupling term, $H_\alpha$, that is present
in the Gell-Mann expansion of $H$, thus \emph{isolating} the term. In
Section \ref{sec:newcouplings} we show that given some term coupling
$k$ qudits, we can simulate new coupling terms that couple fewer than
$k$ qudits.  We also discuss the limitations on this type of
simulation. Section~\ref{sec:universality} examines how we can use a
term that couples $k$ qudits to simulate a coupling between two
qudits.  Finally we prove the main result of the paper: that the only
non-universal class of entangling Hamiltonians is the class of odd
Hamiltonians. This is argued through an exhaustive demonstration that
all $n$-qudit entangling Hamiltonians other than the odd many-qubit
Hamiltonians are indeed universal.

\section{Simple simulations}
\label{sec:simple_sims}

%
In this section, we review some simple Hamiltonian simulation
techniques studied in previous papers~\cite{Dodd02a, Wocjan02a,
  Bennett01a, Leung01a, Dur01a, Nielsen01c, Wocjan02b, Wocjan02c,
  Vidal01b, Vidal01c}, and that will form the basis for our later
results. By a Hamiltonian simulation we mean a sequence of evolutions
due to our system Hamiltonian, $H$, which is assumed fixed,
interleaved with single-qudit unitary operations.  The goal is to
approximate (to arbitrary accuracy) evolution according to some other
Hamiltonian.  If that is possible for some desired Hamiltonian we say
that Hamiltonian can be \emph{simulated}.  The theory of Lie algebras
and Lie groups ensures that the techniques decribed in this section
exhaust the set of possible simulations that can be performed given
some Hamiltonian and single-qudit unitaries.

\subsection{Conjugation by a unitary operator}
\label{susec:conj}

A quantum system with Hamiltonian $H$ evolves in time via the unitary
operation $e^{-iHt}$. Say we are also given the ability to perform
some unitary operation, $U$, and its inverse, $U^\dagger$. Then
performing the sequence of unitary operations $Ue^{-iHt}U^\dagger =
e^{-iUHU^\dagger t}$, we see that we can simulate an evolution
according to the conjugated Hamiltonian $UHU^\dagger$.  In this paper,
as we have given ourselves the ability to perform arbitrary
single-qudit unitaries, we will often conjugate a Hamiltonian by
unitaries of the form $U=U_1\otimes U_2 \otimes \ldots \otimes U_n$.

\subsection{Simulating linear combinations}
\label{subsec:lin_combs}

Suppose we can simulate two different Hamiltonians, $H_1$ and
$H_2$. Then we can simulate the sum of these Hamiltonians, since
$e^{-iH_1
  \Delta}e^{-iH_2 \Delta}\approx e^{-i(H_1 +H_2)\Delta}$ for small
$\Delta$, and with successive evolutions we can simulate the
Hamiltonian $H_1 +H_2$ for an arbitrary time $t$.  Imagine that we
could evolve our system by a whole set of Hamiltonians,
$\mathcal{H}$, and their negatives\footnote{Given that we can
simulate $H$, it turns
  out always to be possible to simulate $-H$, using single-qudit
  unitary operations.  This follows from
  Equation~(\ref{eq:depolarizific}), later in the paper, which shows
  how to express $-H$ as a sum of terms of the form $U H U^\dagger$,
  where $U$ are local unitary operations.  By the methods of
  simulation we've already introduced, it follows that $-H$ can be
  simulated.}.  It follows that we can simulate arbitrary linear
combinations of any of the elements of $\mathcal{H}$.

\subsection{Simulating commutators of Hamiltonians}
\label{subsec:comm_sim} Another simple simulation protocol that
can be performed is the simulation of a commutator of two different
Hamiltonians. This is possible as $e^{-iH_1 \Delta}e^{iH_2
  \Delta}e^{iH_1 \Delta}e^{-iH_2 \Delta}\approx
e^{-i(i[H_1,H_2])\Delta^2}$. So if we can simulate $H_1, H_2$ and
their negations we can simulate the commutator of these Hamiltonians.

\subsection{Simulating Hamiltonians that couple the same qudits}
\label{subsec:maj}

Consider the general expression for a Hamiltonian acting on a system
of qudits in Equations~(\ref{H}) and~(\ref{Halpha}), and recall that
$H_\alpha$ couples a set of qudits $S_\alpha$. We now introduce a
theorem from \cite{Nielsen01c} to show it is possible to use
$H_\alpha$ and single-qudit unitaries to exactly simulate any other
coupling term that couples the set of qudits $S_\alpha$:
\begin{theorem}
\label{thm:maj} Let $A$ and $B$ be any two traceless Hermitian
operators in $d$ dimensions and assume that B$\neq$ 0. There is an
algorithm to find a set of at most $d^2$ unitary operators, $U_n$, and
constants $c_n> 0$ such that:
\begin{equation}
A=\sum_n c_nU_n BU^{\dagger}_n.
\end{equation}
\end{theorem}
Key to proving this theorem is a result from the theory of operator
majorization, Uhlmann's theorem~\cite{Uhlmann71a}. Although we do not
need the theory of majorization in this paper, for the benefit of
readers familiar with majorization, we make the following summary
remarks.  Recall that Uhlmann's theorem tells us that if $P\prec Q$
(that is, $P$ is majorized by $Q$) then $P=\sum_n p_nU_nQ
U_n^\dagger$, for some unitary operators $U_n$ and some $p_n$ that
form a probability distribution.  The proof of Theorem~\ref{thm:maj}
in \cite{Nielsen01c} follows by showing that $A\prec cB$ for some
positive constant $c$.

Any coupling term $H_\alpha$ in $H$ is a tensor product of traceless
terms acting on $S_\alpha$. If we replace $B$ in Theorem~\ref{thm:maj}
by the individual tensor factors appearing in $H_\alpha$, then we see
that we can simulate any $A$ that is a tensor product of traceless
Hermitian operators acting on the same set $S_\alpha$.  This result
will be extremely useful in the remainder of this paper. It tells us
that if we can simulate some coupling $H_\alpha$, we can simulate
every other coupling on the same set of qudits.



\section{Term isolation}
\label{sec:term_isolation}

In Section~\ref{subsec:maj}, we saw that any coupling term,
$H_\alpha$, in the expansion $H=\sum_{\alpha} h_\alpha H_\alpha$
(Equation~(\ref{H})), could be used to simulate any other coupling
term that entangles the same set of qudits.  If we have a Hamiltonian
that is simply a coupling term on a given set of qudits, we can
immediately say a great deal about what can be simulated with that
Hamiltonian. In general we do not have this luxury of interpretation.
Instead, some general Hamiltonian, $H=\sum_\alpha H_\alpha$, has many
different coupling terms that couple many different sets of qudits.
\emph{Term isolation} is a simulation technique that uses $H$ and
single-qudit unitaries to simulate any particular term $H_\alpha$ in
the expansion of $H$ alone.

Term isolation allows us to think about $H$ in a different way,
showing that the ability to simulate $H$ is \emph{equivalent} to the
ability to simulate the coupling terms $\{H_\alpha\}$ individually.
Thus, we can perform our analysis entirely in terms of the set
$\{H_\alpha\}$ and still encapsulate all of the Hamiltonian simulation
properties of $H$. Given that the elements of the set $\{H_\alpha\}$
have a much simpler structure than a general $H$, term isolation is a
powerful tool for analysis.

We now show that term isolation can always be performed. If we
demonstrate that we can use $H$ and single-qudit unitaries to simulate
some $H_\alpha$ coupling an arbitrarily chosen set of qudits, then we
know from Section~\ref{subsec:maj} that it can be used to simulate any
other term coupling the same qudits.

Without loss of generality we may assume that the term being
isolated is of the form
\begin{equation}
\label{eq:keeper} H_\alpha = \bigotimes_{j=1}^{k}
W_{b_j}^{(j)}\otimes I^{\otimes n-k},
\end{equation}
where $k$ is the number of qudits in the set $S_\alpha$.  To see
that there is no loss of generality in assuming this form, note
that we can always relabel the qudits in $S_\alpha$ so that they
are the first $k$ qudits in the system, and any operators
$X_{ab}$ or $Y_{ab}$ in $H_\alpha$ are
equivalent under local unitaries to $W_2$.

Any term in the expansion of $H$, $H_\beta$, that isn't the term $H_\alpha$
that we wish to keep, is different from $H_\alpha$ in at least one of three
ways. Either:
\begin{description}
    \item[\textbf{Case 1:}] $H_\beta$ has terms acting non-trivially on qudits outside of $S_\alpha$, the set of qudits upon which
  $H_\alpha$ acts.

    \item[\textbf{Case 2:}] $H_\beta$ acts on a strict subset of $S_\alpha$.

    \item[\textbf{Case 3:}] $H_\beta$ acts on the same qudits as
      $H_\alpha$ but is a tensor product of different elements of the
      Gell-Mann basis. That is, $H_\alpha\neq H_\beta$, even though
      $H_\beta$ couples the set $S_\alpha$.
\end{description}
Each of these cases identifies a special difference between $H_\alpha$
and $H_\beta$. In the following sections these differences are
exploited to define simulations that remove undesirable terms.

As we have previously stated, every simulation in this section may be
represented as a sequence of linear combinations, commutators and
conjugations by local unitaries. We often denote a sequence of
operations of this type on a Hamiltonian, $H$, by a scripted letter.
For example, in Section~\ref{subsec:depolarizingstuff} we define the
depolarizing channel, which is a linear combination of conjugations by
local unitaries, and write $\mathcal{D}[H]=H_D$ to symbolize the
depolarizing channel acting on $H$, resulting in the simulated
Hamiltonian $H_D$.  The action of $\mathcal{D}$ on $H$ defines a
simulation. We can also compose simulation techniques, so, for
example, in Section~\ref{subsec:moredepolarizingstuff} we define a
simulation $\mathcal{T}[H_D]= H_T$.

\subsection{Case 1}
\label{subsec:depolarizingstuff}

We begin by noting the identity
\begin{equation}
\label{eq:depolarizing} \sum_{U_p} U_pJU_p^{\dagger} = d\text{
}\text{tr}(J)I,
\end{equation}
where $J$ is an operator acting on some qudit of dimension $d$ and the
sum is over all $d^2$ elements of the $d$-dimensional Pauli
group\footnote{The properties of the $d$-dimensional Pauli group were
  extensively studied in~\cite{Gottesman97a}. We will not use any
  further special properties of this group and refer the interested
  reader to~\cite{Gottesman97a} for further information.}, where we
omit repeated summation when two elements in the Pauli group differ
merely by a phase factor.  We note there is a simple extension of
Equation~(\ref{eq:depolarizing}) for multiple-qudit systems,
\begin{eqnarray}
 & & \sum_{U_p^{(j)}}
(U_p^{(1)}\otimes \ldots \otimes U_p^{(n)})J(U_p^{(1)}\otimes \ldots \otimes
U_p^{(n)})^\dagger \nonumber \\
 & = & D\text{ tr}(J)I^{\otimes n},  \label{eq:depolarize1}
\end{eqnarray}
where the superscripts indicate the different qudit systems, of
respective dimension $d^{(j)}$, $D=d^{(1)}...d^{(n)}$ is the dimension
of the combined system, $I$ represents the appropriate identity
operator for each subsystem, and the sum is over conjugations by all
elements of the Pauli group for each qudit, again omitting repeated
sums over elements that are the same up to a phase factor.

We define the simulation $\mathcal{D}[H]=H_D$ to be the
multiple-qudit depolarizing channel acting on the $n-k$ qudits
that aren't coupled by $H_\alpha$,
\begin{eqnarray}
\label{eq:depolarizesim} \mathcal{D}[H] & = & \sum_{U_p^{(j)}}
(U_p^{(k+1)}\otimes...\otimes U_p^{(n)})H
(U_p^{(k+1)}\otimes...\otimes U_p^{(n)})^\dagger \nonumber
\\ & = &H_D.
\end{eqnarray}
$H_\alpha$ acts on the first $k$ qudits of an $n$-qudit system, that
is, the set $S_\alpha$. If we examine the simulated Hamiltonian,
$H_D$, we find from Equation~(\ref{eq:depolarize1}) that any terms
$H_\beta$ in $H$ that act non-trivially on qudits outside the set
$S_\alpha$ are eliminated.  The simulation leaves the coupling term
$H_\alpha$ unchanged except for an unimportant positive scaling
factor.  Thus we have removed all the Case~1 terms $H_\beta$ from the
Hamiltonian, and need only consider the remaining Case~2 and Case~3
terms.

\subsection{Case 2}
\label{subsec:moredepolarizingstuff}
The Hamiltonian $H_D$ is a linear combination of terms that couple the
set of qudits $S_\alpha$ or some subset of $S_\alpha$.  It turns out
that we can use another extension of Equation~(\ref{eq:depolarizing})
to simulate a Hamiltonian, $H_T$, that only has terms that couple the
set $S_\alpha$. In Equation~(\ref{eq:depolarizing}), if $J$ is a
traceless operator we find that the right hand side of the equation is
zero. Noting that $I$ is an element of the Pauli group, we find
\begin{equation}
\label{eq:depolarizific} \sum_{U_p\neq I} U_p J U_p^\dagger = -J,
\end{equation}
which always holds for traceless $J$. Using single-qudit unitaries
from the Pauli group we consider the following summation,
\begin{equation}
\label{eq:depdouble} \sum_{U_p^{(1)} \neq I, U_p^{(2)}\neq I}
(U_p^{(1)}\otimes U_p^{(2)}) (J^{(1)}\otimes
J^{(2)})(U_p^{(1)}\otimes U_p^{(2)})^\dagger.
\end{equation}
If $J^{(1)}$ and $J^{(2)}$ are traceless, this expression is equal to
$J^{(1)}\otimes J^{(2)}$. If $J^{(2)}$ is traceless and $J^{(1)}$ is
the identity, this expression is equal to $-[(d^{(1)})^2-1] I\otimes
J^{(2)}$. With this in mind we define a simulation:
\begin{eqnarray}
\mathcal{T}^{(j)}[H]& \equiv & ((d^{(j)})^2-1)H \nonumber \\
& & + \sum_{U_p^{(1)}, U_p^{(j)}\neq I} (U_p^{(1)}\otimes
U_p^{(j)}) H(U_p^{(1)}\otimes U_p^{(j)})^\dagger. \nonumber\\
\end{eqnarray}
Performing $\mathcal{T}^{(j)}$ for $j=2,...,k$, only terms that
couple the same qudits as $H_\alpha$ are not eliminated. So
performing the following sequence of simulations,
\begin{equation}
\mathcal{T}[H_D] =
\mathcal{T}^{(k)}[\mathcal{T}^{(k-1)}[...[\mathcal{T}^{(2)}[H_D]]...]]=H_T
\end{equation}
the simulated Hamiltonian, $H_T$, is a linear combination of terms
that couple the same qudits as $H_\alpha$.

\subsection{Case 3}
We have shown how to simulate a Hamiltonian $H_T$ that only contains
terms which couple the same qudits as $H_\alpha$.  To eliminate the
remaining terms we define the following operators that are both
unitary and Hermitian,
\begin{equation}
Z_a \equiv I-2|a\rangle \langle a| = \sum_{j=1}^d |j\rangle\langle j|-
2|a\rangle\langle a|.
\end{equation}
Notice that the $Z_a$ operators commute with each of the
Cartan subalgebra elements, $W_m$, in Equation~(\ref{eq:W_m}).
Hence, each of the $Z_a$ will also commute with
$H_\alpha$ as it is a tensor product of elements of the Cartan
subalgebra. Further notice that $Z_a$ anti-commutes with
$X_{lm}$ and $Y_{lm}$ if $a=l$ or $a=m$ and
commutes otherwise. We can use this fact to define a simulation
that eliminates terms with $X_{lm}$ and $Y_{lm}$
operators present in $H_T$. We define a simulation
\begin{equation}
\mathcal{Z}_a^{(j)}[H] = H+Z_a^{(j)} H
Z_a^{(j)},
\end{equation}
where the superscript $j$ indicates a $Z_a$ operator
acting on the $j$th qudit, with identities acting elsewhere. If
there exists any term with an $X_{lm}$ or
$Y_{lm}$ operator on the $j$th qudit, and such that $a=l$
or $a=m$, then this term will be eliminated from $H_T$ by the
simulation $\mathcal{Z}_a^{(j)}[H_T]$. Expanding on this idea we
can eliminate every term on the $j$th qudit that has the form
$X_{lm}$ or $Y_{lm}$ by performing the following
simulation:
\begin{equation}
\mathcal{Z}^{(j)}[H_T] \equiv
\mathcal{Z}^{(j)}_d[\mathcal{Z}^{(j)}_{(d-1)}[...[\mathcal{Z}_1^{(j)}[H_T]]...]]
\end{equation}
where $d$ is the dimension of the $j$th qudit. The effect of this
simulation on $H_\alpha$ is simply to rescale it. Now, if we perform
the simulation $\mathcal{Z}^{(j)}$ for each qudit in $S_\alpha$,
\begin{equation}
\mathcal{Z}[H_T] =
\mathcal{Z}^{(k)}[\mathcal{Z}^{(k-1)}[...[\mathcal{Z}^{(1)}[H_T]]...]]=H_Z,
\end{equation}
all that remains in the newly simulated Hamiltonian, $H_Z$, is a
linear combination of terms that commute with the Cartan
subalgebra elements. We have now simulated a Hamiltonian with no
$X$- and $Y$-type terms.

$H_Z$ is a linear combination of terms that are tensor products of
operators from the Cartan subalgebra. Consider the unitary
representation, $P^{(j)}(\pi)$, of the permutation group $S_{b_j-1}$
that permutes the elements of the diagonal basis of the Cartan
subalgebra, $|a\rangle$, for $a = 1,\ldots,b_j-1$ on the $j$th qudit.
When $a \geq b_j$ we have $P^{(j)}(\pi)W_a P^{(j)\dagger}(\pi) = W_a$.
When $a<b_j$, we find that the effect of conjugating $W_a$ by a
permutation operation is to shift around the diagonal elements of
$W_a$. Now, we can eliminate any terms in $H_Z$ that contain an
operator $W_a^{(j)}$ with $a<b_j$ by performing the simulation
\begin{equation}
\mathcal{P}^{(j)}[H_Z] = \sum_{\pi\in S_{b-1}}
P^{(j)}H_Z P^{(j)\dagger}.
\end{equation}
This works because $W_a^{(j)}$ is a diagonal, traceless operator and
the permutation, $\mathcal{P}^{(j)}$, distributes each of the diagonal
elements of $W_a^{(j)}$ equally. The effect of $\mathcal{P}^{(j)}$ on
terms $W_a^{(j)}$ acting on the $j$th qudit and with $a\geq b_j$ is to
simply scale them by a factor of $(b_j-1)!$.  Performing the following
simulation,
\begin{equation}
\mathcal{P}[H_Z] =
\mathcal{P}^{(k)}[\mathcal{P}^{(k-1)}[...[\mathcal{P}^{(1)}[H_Z]]...]]=H_P
\end{equation}
we produce a Hamiltonian $H_P$ that is a linear combination of
terms that couple the same qudits as $H_\alpha$ and are tensor
products of operators $W_a$ with $a\geq b_j$.

In Section~\ref{subsec:comm_sim} we pointed out that it is possible to
simulate a Hamiltonian proportional to the commutator of two
Hamiltonians that are both simulatable. Now, we note that the
commutator $-i[W_a^{(j)},X_{b_j-1 \, b_j}]=0$ if $a>b_j$. If $a=b_j$
we find $-i[W_{b_j}^{(j)}, X_{b_j-1 \, b_j}]=
\frac{\sqrt{b_j}}{\sqrt{b_j-1}} Y_{b_j-1 \, b_j}$. We can make use of
this distinction to find a way to remove the unwanted terms from
$H_P$. We define the simulation
\begin{equation}
\mathcal{X}^{(j)}[H] \equiv -i[H,X^{(j)}_{b_j-1 \, b_j}].
\end{equation}
Then if we perform the following sequence of simulations,
\begin{equation}
\mathcal{X}[H_P] =
\mathcal{X}^{(k)}[\mathcal{X}^{(k-1)}[...[\mathcal{X}^{(1)}[H_P]]...]]=H_X
\end{equation}
we find that $H_X = \left( \otimes_{j=1}^k Y_{b_j-1,b_j}
\right) \otimes I^{\otimes n-k}$, up to some unimportant but non-zero
constant multiple.  We have now simulated a single coupling term that
couples the same qudits as $H_\alpha$. Recall in
Section~\ref{subsec:maj} we noted that a coupling term can be used
with single-qudit unitaries to simulate any other term coupling the
same set of qudits. So, we can use $H_X$ and single-qudit unitaries to
simulate $H_\alpha$, the desired term. Thus we have demonstrated that
it is possible to isolate $H_\alpha$ from $H$.

\section{Simulating new coupling terms}
\label{sec:newcouplings}

Term isolation shows that the ability to simulate a Hamiltonian $H
=\sum_\alpha h_\alpha H_\alpha$ is equivalent to the ability to
simulate the set of coupling Hamiltonians, $\{H_\alpha\}$, given
single-qudit unitary operations.  Additionally, we learnt in
Section~\ref{subsec:maj} that given $H_\alpha$ and single-qudit
unitaries we can simulate any coupling term that couples the same
qudits as $H_\alpha$.  So far we have not presented any way of
simulating some coupling term that couples a different set of qudits
than any of the terms in the set $\{H_\alpha\}$.  In this section we
will take a key step towards a proof of universality, showing how to
use single-qudit unitaries and a term $H_\alpha$ coupling $k$ qudits
in order to simulate a term that couples $k-1$ qudits.

\subsection{Evaluation of commutators}
\label{subsec:commexample}
%
%

In \cite{Bremner03a} it was shown that if $H_\alpha$ coupled qubits,
its capacity to simulate other coupling terms depended on the number
of qubits that it coupled. More specifically, it was shown that if
$H_\alpha$ coupled $k$ qubits and $k$ was an odd number, then
$H_\alpha$ couldn't be used with single-qubit unitaries to simulate a
coupling term that coupled $k-1$ qubits.  One way of seeing why this
is true is to examine the commutator of two Hamiltonians, $[H_\alpha,
H_\beta]$, that couple the same set of qubits $S_\alpha$. It is easy
to show that the commutator $[H_\alpha, H_\beta]\neq 0$ if and only if
there are an odd number of locations in $S_\alpha$ where $H_\alpha$
and $H_\beta$ differ.  {} From this restriction it is possible to
prove, as was done in \cite{Bremner03a}, that coupling terms coupling
an odd number of qubits can only ever simulate other Hamiltonians that
have odd couplings.

What is different when not all the systems are qubits?  The purpose of
this subsection is to investigate the commutator of two specially
chosen couplings $H_\alpha$ and $H_\beta$ that couple the same set of
qudits, $S_\alpha$.  In the case of qubits, it is not difficult to
convince oneself that when $S_\alpha$ contains an \emph{even} number
of qubits, the commutator $[H_\alpha,H_\beta]$ is either zero, or else
couples a set of qubits that is a \emph{strict} subset of the original
set $S_\alpha$.  We will show by an explicit calculation that when one
or more of the systems is not a qubit, it is possible to choose
$H_\alpha$ and $H_\beta$ so that the commutator $[H_\alpha,H_\beta]$
contains terms coupling the entire set $S_\alpha$.  Remarkably, we
will see in the remainder of the paper that this is the key fact that
simplifies the study of universality when not all the systems are
qubits.

We begin by choosing $H_\alpha = \bigotimes_{j=1}^{k} X_{ab}^{(j)}$
and $H_\beta = \bigotimes_{j=1}^{k} X_{ab'}^{(j)}$ where for all $j$
we set $b\neq b'$.  (We assume initially that all systems are of
dimension $3$ or greater.) Given these forms for $H_\alpha$ and
$H_\beta$, what does $[H_\alpha,H_\beta]$ look like? We find
\begin{eqnarray}
[H_\alpha,H_\beta]& = & \bigotimes_{j=1}^k
\frac{1}{2\sqrt{2}}(X_{bb'}^{(j)}+iY_{bb'}^{(j)})
\nonumber
\\ &  &-\bigotimes_{j=1}^k
\frac{1}{2\sqrt{2}}(X_{bb'}^{(j)}-iY_{bb'}^{(j)}).
\end{eqnarray}
This expression contains Hermitian and skew-Hermitian terms. Upon
expansion of the above expression we find that all of the Hermitian
terms sum to zero, leaving only a sum of skew-Hermitian terms
remaining. These terms correspond to a sum of tensor product terms
containing odd numbers of $Y_{bb'}$ terms.  All of the terms couple
the entire set $S_\alpha$.  It is easy to verify that this sum is
always non-zero, simply by inspection of the coefficients of the
relevant terms.

So far we have only considered the case where we could choose to
simulate $H_\alpha$ and $H_\beta$ for $X_{ab}^{(j)}$ and
$X_{ab'}^{(j)}$, $b\neq b'$. We can only do this when each subsystem
has dimension $d>2$. If we have subsystems where $d=2$, the situation
changes slightly, but the results are similar, provided not all of the
subsystems are qubits.

For every $j$ where the qudit has dimension $d>2$ we choose
$H_{\alpha}^{(j)} = X_{ab}^{(j)}$ and $H_{\beta}^{(j)}=X_{ab'}^{(j)}$
with $b\neq b'$. For every $j$ where the qudit has dimension $d = 2$,
we choose $H_{\alpha}^{(j)}=X$, and $H_{\beta}^{(j)}=Y$.  Provided
$H_\alpha$ and $H_\beta$ do not couple qubits exclusively, a
straightforward calculation along lines similar to that already done
shows that $[H_\alpha,H_\beta]$ is a non-zero sum of terms, each of
which is skew-Hermitian and couples all $k$ qudits.  The only subtlety
in the calculation is the need to analyse separately the cases where
there are an \emph{even} number of qubits in the set $S_\alpha$, which
gives rise to a commutator which is a non-zero sum of tensor product
terms containing an odd number of $Y_{bb'}$ terms, and the case where
there are an \emph{odd} number of qubits in the set $S_\alpha$, which
gives rise to a commutator which is a non-zero sum of tensor product
terms containing an even number of $Y_{bb'}$ terms.

\subsection{Simulating identity operators}

Given some term, $H_\alpha$, coupling a set of qudits $S_\alpha$, we
show how the results on commutators just obtained allow us to simulate
other coupling term that couples a subset of $S_\alpha$ with just one
qudit removed. More precisely:

\begin{lemma}
\label{lemma:Imaking} Given the ability to evolve via $H_\alpha =
\bigotimes_{j=1}^n H_{\alpha}^{(j)}$, which couples $k$ qudits,
and local unitary operations, it is possible to simulate $H'$
such that
\begin{equation}
\label{eq:Iwithcomm} H'= I\otimes H_\gamma,
\end{equation}
provided $H_\alpha$ does not couple qubits exclusively.  The coupling
term $H_\gamma$ may couple any $k-1$ qudit subset of $S_\alpha$,
subject to the constraint that the subset not be qubits exclusively.
\end{lemma}

\textbf{Proof:} Given $H_\alpha$ we can simulate any other coupling
term, $H_\beta = \otimes_{j=1}^n H_{\beta}^{(j)}$, that acts
non-trivially on the same set of $k$ qudits, $S_\alpha$.  We label the
qudits so that $S_\alpha$ consists of qudits $1,\ldots,k$, and so that
our goal is to simulate a coupling on qudits $2,\ldots,k$, i.e., the
goal is to remove qudit $1$.  To this end, we choose $H_{\beta}^{(1)}$
so that $H_{\alpha}^{(1)}=H_{\beta}^{(1)}$.  Note that, by assumption,
the set $2,\ldots,k$ does not contain qubits exclusively.  Evaluating
the commutator, we find:
\begin{equation}
i[H_\alpha,H_\beta]= i
(H_{\alpha}^{(1)})^{2}\otimes\left[\bigotimes_{j=2}^nH_{\alpha}^{(j)},\bigotimes_{j=2}^nH_{\beta}^{(j)}\right].
\end{equation}
Setting $N\equiv \bigotimes_{j=2}^n H_{\alpha}^{(j)}, N' \equiv
\bigotimes_{j=2}^{n}H_{\beta}^{(j)}$, and applying
Equation~({\ref{eq:depolarizing}) to the first qudit, we see that it
  is possible to simulate
\begin{equation}
H'=i I\otimes[N,N'].
\end{equation}
Finally, we note that as $N$ and $N'$ don't act exclusively on qubits,
our earlier results on commutators show that we can ensure that
$[N,N']$ is a non-zero linear combination of terms that couple
$S_\alpha$, less the first qudit.  Term isolation allows us to
simulate one of the coupling terms in $[N,N']$ alone, i.e., $H''=
I\otimes H_\gamma$, as required. $\Box$

\section{Universality}
\label{sec:universality}
%
%
Theorem~\ref{thm:2local} stated that if a set of qudits is
\emph{connected} by a Hamiltonian, $H$, with two-body interactions,
then evolutions by $H$ and single-qudit unitaries form a universal set
of operations on that set of qudits~\cite{Wocjan02c}. A set of 2-qudit
coupling terms connecting the same set of qudits is also universal as
they can simulate a two-body Hamiltonian on the set of qudits. We
prove in this section the main result of this paper: that a generic
Hamiltonian, $H$, entangling a set of qudits can simulate a set of
2-qudit coupling terms connecting the qudits, and is thus universal.
The only exception to this rule is the case where $H$ is a sum of odd
coupling terms, as discussed in~\cite{Bremner03a}, and summarized in
Theorems~\ref{thm:nqubituniv} and~\ref{thm:oddhams} in the present
paper.

We begin by proving Theorem \ref{thm:univ1}, which shows that a
coupling term, $H_\alpha$, that couples a set of $k$ qudits,
$S_\alpha$, can be used to simulate a set of 2-qudit couplings that
connect the set $S_\alpha$. This implies that $H_\alpha$ and
single-qudit unitaries are a universal set on the qudits $S_\alpha$.
We conclude with Theorem~\ref{thm:univtastic}, showing that an
arbitrary entangling Hamiltonian on $n$ qudits is universal for the
qudits it entangles.

\subsection{Theorem~\ref{thm:univ1}: Using a term coupling many
qudits to simulate a term coupling two qudits.}


\begin{theorem}
\label{thm:univ1} Suppose $H_\alpha = \bigotimes_{j=1}^n
H_{\alpha}^{(j)}$ couples $k$ qudits.  Then $H_\alpha$ and
single-qudit unitary operations can be used to simulate a set of
two-qudit couplings connecting every qudit coupled by $H_\alpha$,
provided $H_\alpha$ does not couple qubits exclusively, and $k>1$.
Thus $H_\alpha$ and single-qudit unitaries are universal on the set of
qudits coupled by $H_\alpha$.
\end{theorem}

\textbf{Proof:} Without loss of generality we may label the systems so
that $H_\alpha$ couples systems $1$ through $k$, and system $1$ is not
a qubit.  Fix $j$ in the range $2$ through $k$.  Applying
Lemma~\ref{lemma:Imaking} repeatedly, we see that we can simulate a
Hamiltonian coupling system $1$ and system $j$.  It follows that
$H_\alpha$ and single-qudit unitaries are universal on the set of
qudits coupled by $H_\alpha$. $\Box$


\subsection{Theorem \ref{thm:univtastic}: Which Hamiltonians are universal?}

With Theorem~\ref{thm:univ1} in mind, we now prove that the only
non-universal set of entangling Hamiltonians is the set of odd
Hamiltonians acting on qubits alone.

\begin{theorem}
\label{thm:univtastic} Single-qudit unitary operations, and
evolutions via a Hamiltonian, $H$, that connects a set of $n$
qudits, is a universal set of operations on those $n$ qudits if
and only if $H$ is not an odd Hamiltonian acting on qubits alone.
\end{theorem}

\textbf{Proof:} The forward implication follows from
Theorem~\ref{thm:nqubituniv}, as does the reverse implication when all
systems are qubits.  Thus, all that needs proof is the reverse
implication in the case when $H$ is an entangling Hamiltonian that
does not act exclusively on qubits.  We will show how to construct a
set of two-body couplings that connect all $n$ qudits.

To construct this set, begin by picking a system that is not a qubit,
and label it system $1$.  We will explain how to construct a set, $S$,
of systems to which $1$ can be coupled via a two-body interaction.  We
begin by setting $S = \{ 1 \}$, and aim to add in other systems that
can be coupled to $1$ via two-body interactions.  Our strategy is to
show that provided $S$ is not yet maximal, i.e., does not yet contain
all $n$ qudits, then it is always possible to add an extra qudit into
$S$.

To see this, suppose $S$ is not yet maximal.  Then it is always
possible to pick a qudit $j$ inside $S$ and a qudit $k$ outside of $S$
such that $H$ contains a coupling term $H_{jk}$ which couples systems
$j$ and $k$.  (Other systems may also be coupled by $H_{jk}$.)  In the
case when either $j$ or $k$ is not a qubit, Theorem~\ref{thm:univ1}
shows that a term coupling just $j$ and $k$ may be simulated.
Theorem~\ref{thm:2local} implies that we can also simulate a term
coupling system $1$ and $k$, and so system $k$ may be added to $S$.

The other possible case is when $j$ and $k$ are both qubits.  In this
case, suppose without loss of generality that $H_{jk}$ has the form
$X^{(j)} \otimes X^{(k)} \otimes \ldots$, where the superscripts label
the systems.  We may also simulate the coupling $X^{(1)}_{12} \otimes
Z^{(j)}$, since system $j$ is in $S$.  Taking the commutator of these
two couplings, we see that we may simulate couplings of the form
$X^{(1)}_{12} \otimes Y^{(j)} \otimes X^{(k)} \otimes \ldots$.
Applying Theorem~\ref{thm:univ1}, we see that it is possible to
simulate a two-body coupling between system $1$ and $k$, and thus
system $k$ may be added to $S$. $\Box$

\section{Conclusion}

We have demonstrated that many-qudit Hamiltonians combined with
local unitary operations are always universal for simulation on
any connected set of subsystems upon which the interactions act
nontrivially, \emph{provided} that Hamiltonian is not an odd
Hamiltonian acting on qubits.  This result is rather intriguing
and elegant, especially in the light of the general lack of broad
results for many-body (as opposed to two-body) problems in quantum
information science. In the study of pure state bipartite
entangled states, for example, a single unit of currency, the
maximally entangled state, has been identified and the fungible
nature of this currency has been established. On the other hand, a
similar currency and set of fungible transformations has not been
identified for systems consisting of more than two parties.  Given
this difficulty in understanding the structure of quantum states,
it is quite remarkable that, with the exception of odd entangling
Hamiltonians, all of the different many-qudit interactions are
equivalent.  Even in the case of odd entangling Hamiltonians,
universal simulation can be achieved using an encoding which
wastes only a single extra qubit of space \cite{Bremner03a}.  Thus
there is a real sense in which, for simulation, all interactions
have been created equal.

Part of the simplicity of our result stems from our focus on
universality for simulation as opposed to universality for quantum
computation, which requires that issues of \emph{efficiency} be taken
into account.  When one adds the requirement of efficiency of
simulation, then problems of universality become much more difficult:
indeed this is perhaps one of the fundamental problems in the study of
the computational complexity of quantum circuits.  A well-developed
theory of efficient simulation is a task of great importance and,
judging from the difficulties encountered in proving lower bounds for
problems in classical circuit complexity, this task is probably an
immensely difficult problem. This paper can be seen, however, as a
necessary precursor to any attempt to advance this program.

\acknowledgements

We thank Jennifer Dodd, Henry Haselgrove and Andrew Hines checking
this manuscript and for helpful discussions. This work was
supported in part by the National Science Foundation under Grant.
No. EIA-0086038.

\end{document}